\newcommand{\beq}{\begin{equation}}
\newcommand{\eeq}{\end{equation}}
\newcommand{\beeq}{\begin{eqnarray}}
\newcommand{\eeeq}{\end{eqnarray}}

\documentstyle[12pt,epsfig] {article}

\begin{document}

\begin{center}\bf
HEAVY STATES OF HADRON  STRING  AS  WEAK INTERACTING  
MASSIVE  PARTICLES  FOR DARK  MATTER  

\bigskip
V. A.  Kudryavtsev  \\  
    Petersburg  Nuclear Physics Institute\\
\bigskip
A b s t r a c t
\end{center}

   Massive states ( of order 10 Gev and more )   of hadron string  ( with scale $ \alpha ^\prime \approx 1 Gev^{-2} $)
can  have very small coupling constants with usual baryons of the Universe.
 Corresponding  mean times for them are found  to be of order
and even more  than  the age of the  Universe.   They  are proposed as  possible
 candidates  of weak  interacting  massive  particles  (WIMP)  for Dark  Matter.

\newpage

            String models  (previously dual resonance models) \cite{1,2} had appeared as 
 a possible way  to describe  strong  interactions at low and intermediate energies.
Presently phenomelogical status of such appproach seems to be even more impressive  than fourty years ago
since we have now stringlike spectrum of hadron states including not only leading Regge trajectories but  and
second and third daughter Regge trajectories for this spectrum \cite{3,4} up to spins equal to be
to 2,3 and even up to 4 for mesons or 11/2 for baryons .  However before  we  had not
consistent realistic string amplitudes for hadrons without negative norm states in physical spectrum and with
 intercept of leading meson ($\rho$) trajectory to be equal to one half \cite {5}.  It turns out to be possible 
to build  a composite string model to be compatible with these requirements \cite{6}. This model  gives 
 realisic description of the hadron spectrum and brings to  correct interaction of $\pi$ -mesons  which satisfies
the Adler-Weinberg condition for soft $\pi$ -mesons \cite{7}.

       All  relativistic quantum string  models  lead to the exponential growth of the number  of string states with mass m of
these states.  Before the string approach  R.Hagedorn had obtained such growth  in 1965 for the statistical bootstrap model  \cite{8}
and had discovered  some correspondence of this behaviour of the number of states  with  experimental data.  According  
to  Hagedorn and  to string approach \cite{9}  this  dependence of the number of hadron states per unit mass  f(m)   is expressed
 asymptotically  by the following form:
\begin{eqnarray}
  f(m)\approx A m^{-\gamma}\exp{\frac{m}{m_0}}
\end{eqnarray}
In the Hagedorn model  ${\gamma}=\frac5{2}; {m_0}\approx 0,16 Gev$.
For string approach  these parameters depend on a effective dimension d, 
 for usual classical string
\begin{eqnarray}
m_0=\frac1{\pi}\sqrt{\frac3{2d \alpha ^\prime }}.
\end{eqnarray}
In  our case of the hadron string model   we  have in critical case:\\
  $d=d_{bos}+ \frac{d_{ferm}}{2}= 8+4=12;  {m_0}\approx  0,12 Gev$;
$ \alpha ^\prime \approx (0,8 -0,9) Gev^{-2} $.
      
      Let us consider contributions of these heavy string states to simple  four-point hadron amplitudes $A+B \rightarrow A+B$ : \\
$\pi+\pi \rightarrow\pi+\pi; \pi+N  \rightarrow \pi+N;
 N+N \rightarrow N+N $  and so on.
    The  pole  contribution for given mass m  contains  the sum of  all  these string constituents  and 
 we can see that this exponential growth of the number of states with mass m  leads to the exponential 
 fall of mean partial widths  $\Gamma_i$  and  corresponding coupling constants .  
  Namely
\begin{eqnarray}
\sum_i \Gamma_i=\Gamma; \\ \nonumber
 \overline{\Gamma}_i=\frac{\Gamma}{ f(m)};\\  \nonumber
 \overline{\Gamma}_i \approx \Gamma  m^{\gamma}\exp{(-\frac{m}{m_0})}
\end{eqnarray}
 The total width $\Gamma$  increases  with mass m not more than  some power of m.
So  for most of these  string states with high masses  we obtain 
very small widths $\Gamma_i $ and hence  highly significant mean times.
      Taking $ \Gamma \approx 1 Gev$  we have
\begin{eqnarray}
   \overline {\tau_i} \approx
 \frac 1{ \overline{\Gamma}_i}\approx \frac {f(m)}{\Gamma}
\approx 10^{-23}\exp{\frac{m}{m_0}}\approx  10^{-23}\exp{8,3m}.
\end{eqnarray}
Since the age of the Universe $ t_0 \approx 13,7 Gyr\approx 4,3 \times 10^{17}s$  we obtain 
$ \overline {\tau_i}> t_0 $  for $ m>11 Gev $.

      So the same heavy states with masses $m>11 Gev $ can be generated in the early Universe
 at $ T>10^{14} K $ and $ t < 10^{-8}s$ from the Big Bang and can survive partly 
up to our time. Certainly we propose that they annihilate  and decay significantly at subsequent times $ t > 10^{-8}s$ 
and  but some part of them ( $10^{-9}$  as for usual baryon matter or maybe much  less than one)   lives through  for $t>10^{-4}$.

 Estimates for cross sections of interactions of these states $A_{dm}$ of mass m with usual hadrons $A_H$ give 
 from the same considerations as above
\begin{eqnarray}
\sigma (A^i_{dm}A_H  )\approx \frac {\sigma (A_H A_H')}{f(m)}\approx\sigma _H (\exp{-8,3m}) 
\end{eqnarray}
It does  not contradict  experimental data  in searches of dark matter to be obtained recently  \cite{10}. 
    
      I would like to thank  participants of joint HEPD-TPD seminar of PNPI
 for discussions of this work.

\end{document}